\documentclass[prl,superscriptaddress,twocolumn]{revtex4}
\usepackage{enumerate}
\usepackage{amsfonts,amssymb,amsmath}
\usepackage[]{graphics,graphicx,epsfig}
\usepackage{amsthm}

\usepackage{graphicx}
\usepackage{dcolumn}
\usepackage{natbib}
\usepackage{color}
\usepackage{multirow}

\def\identity{\leavevmode\hbox{\small1\kern-3.8pt\normalsize1}}

\newtheorem{propo}{Proposition}
\newcommand{\be}{\begin{eqnarray}}
\newcommand{\ee}{\end{eqnarray}}
\newcommand{\bpr}{\begin{propo}}
\newcommand{\epr}{\end{propo}}
\newcommand{\bpf}{\begin{proof}}
\newcommand{\epf}{\end{proof}}
\newcommand{\ket}[1]{\left | #1 \right\rangle}
\newcommand{\bra}[1]{\left \langle #1 \right |}

\renewcommand{\epsilon}{\varepsilon}
\begin{document}

%%%%%%%%%%%%%%%%%%%%%%%%%%%%%%%%%%%%%%%%%%%%%%%%%%%%%%%%%%%%%%%%%%%%
\title{Genuine multipartite indistinguishability and its detection via generalized Hong-Ou-Mandel effect}

\author{Marcin Karczewski}   
\affiliation{Faculty of Physics, Adam Mickiewicz University, Umultowska 85, 61-614 Pozna\'n, Poland}

\author{Robert Pisarczyk}   
\affiliation{Mathematical Institute, University of Oxford, Woodstock Road, Oxford OX2 6GG, U.K.}
\affiliation{Centre for Quantum Technologies, National University of Singapore, 3 Science Drive 2, 117543 Singapore, Singapore}

\author{Pawe\l{} Kurzy\'nski}   \email{pawel.kurzynski@amu.edu.pl}   
\affiliation{Faculty of Physics, Adam Mickiewicz University, Umultowska 85, 61-614 Pozna\'n, Poland}
\affiliation{Centre for Quantum Technologies, National University of Singapore, 3 Science Drive 2, 117543 Singapore, Singapore}

\date{\today}

%%%%%%%%%%%%%%%%%%%%%%%%%%%%%%%%%%%%%%%%%%%%%%%%%%%%%%%%%%%%%%%%%%%%

\begin{abstract}

The question whether two indistinguishable particles are bosons or fermions can be answered by observing the Hong-Ou-Mandel effect on a beam splitter. However, already for three particles one can consider symmetries that are neither bosonic nor fermionic. In this work, we describe a simple method of identifying them experimentally and propose a measure of a genuine multipartite indistinguishability.

\end{abstract}

\maketitle

%%%%%%%%%%%%%%%%%%%%%%%%%%%%%%%%%%%%%%%%%%%%%%%%%%

{\it Introduction.} The concept of indistinguishability is rooted in the permutation symmetry. In essence, particles are indistinguishable if the underlying state remains unchanged after a permutation of their labels. This means that there are only two types of indistinguishable pairs of  particles, bosons and fermions, since there are only two possibilities for such a system not to be affected by a transposition  \cite{note}. However, for more than two particles the permutation symmetry allows for new possibilities, known as parastatistics \cite{hartle69,bach97,polychronakos99}.   

They emerge from the fact that in the multipartite case there are many  inequivalent permutations of the particles' labels.  Usually, the notion of indistinguishability is defined with respect to all of them, but one can also consider it regarding only a small subset. Here, we focus on cyclic permutations and show that they lead to the concept of genuine multipartite indistinguishability. We further show that genuine multipartite indistinguishable states exhibit unique dynamical properties on symmetric multiports.

This feature stems from a non-trivial multipartite interference effect \cite{Dittel18} that connects the system's symmetry with suppressed probabilities of certain outcomes on multiports. In particular, we show that just as the Hong-Ou-Mandel effect \cite{HOM} can be used to differentiate between two bosons and two fermions, the generalized suppression laws enable the detection and characterization of genuine multipartite indistinguishability.  Moreover, they provide its clear operational measure.

There are three main motivations behind our research. Firstly, indistinguishability can be considered  a potential quantum resource \cite{franco18,blasiak18,GPT, monogamy}  related to, yet different from entanglement. Since the resource theory of entanglement \cite{Horodecki} had a large impact on many fields of quantum science, we believe that a parallel development of the resource theory of indistinguishability can also prove to be a worthy endeavor. To lay its foundations, we discuss   indistinguishability outside of a standard bipartite setting.  In particular, we provide a rigorous definition of a genuine multipartite indistinguishability and propose its  experimentally feasible  measure.  

Secondly, indistinguishability is studied as a necessary condition for the multipartite interference \cite{Brod18}.  From this perspective, the states we investigate are particularly interesting, as they consist of particles that share some form of global indistinguishability. This leads to peculiar interference effects, that might find some applications in quantum information protocols. For example, a form of bunching in which not all the particles tend to group together could be exploited to tweak the evolution of multipartite quantum walks \cite{omar06,sciarrino12}. Moreover, multipartite permutation symmetric states can be used in secret sharing scenarios \cite{cabello02,wang16}.

Finally, we are also interested in the dynamical properties of the parastatistical states in the context of genuine multipartite entanglement. Note, that the evolution of correlated non-interacting classical particles can be considered a mixture of independent evolutions, studied particle by particle. On the other hand, due to non-classical correlations  a collection of non-interacting quantum particles may evolve in a fundamentally multipartite way. An example of such behaviour is the bunching of bosons and the antibunching of fermions. Here, we study such evolutions for various parastatistical states. We observe non-trivial patterns in the particle-count statistics which we interpret as a signature of genuine multipartite entanglement.
%%%%%%%%%%%%%%%%%%%%%%%%%%%%%%%%%%%%%%%%%%%%%%%%%%%%%%%%%%%%%%%%%%%%

%\section{Preliminaries}

{\it Preliminaries.} Consider a quantum system made of $n$ particles. For simplicity we assume that it can be described by a pure state

\begin{equation}
|\psi\rangle = \sum_{k_1,\ldots,k_n}\alpha_{k_1,\ldots,k_n}|k_1 \ldots k_n\rangle,
\end{equation}
where $k_i$ denotes the state of the $i^{\text{th}}$ particle.

The goal is to determine if these particles can be considered indistinguishable. As already mentioned in the introduction, this property is connected with permutation invariance. For instance, we say that  two particles, $i$ and $j$, are indistinguishable if the swap of their labels
\begin{eqnarray}
\Pi_{ij}|\psi\rangle &=& \Pi_{ij}\sum_{k_1,\ldots,k_n}\alpha_{k_1,\ldots,k_n}|k_1 \dots k_i \dots k_j \dots k_n\rangle \nonumber \\
&=& \sum_{k_1,\ldots,k_n}\alpha_{k_1,\ldots,k_n}|k_1 \dots k_j \dots k_i \dots k_n\rangle 
\end{eqnarray}
does not change the state of the system, i.e. $|\langle \psi |\Pi_{jk} |\psi\rangle |^2=1$.

The above condition leads to a natural measure of bipartite indistinguishability 
\begin{equation}\label{bim}
{\cal I}_{ij} \equiv |\langle\psi| \Pi_{ij}|\psi\rangle|^2.
\end{equation}
Note that $0 \leq {\cal I}_{ij} \leq 1$. The value 1 is attainable if particles $i$ and $j$ are indistinguishable, i.e., they are in a symmetric (bosonic) or antisymmetric (fermionic) state. On the other hand, the value 0 occurs if the state after the permutation is orthogonal to the initial one, which implies perfect distinguishability. In other words, here we do not ask whether the particles are bosons or fermions but rather whether they are indistinguishable or not.

As the swap of particle labels is not a physical operation, one may question if there exists an experimentally feasible method to  measure ${\cal I}_{ij} $. However, it turns out that this value can be easily obtained by investigating the interference phenomenon known as the Hong-Ou-Mandel effect \cite{HOM}. When two particles are cast into different input ports of a symmetric beam splitter, they leave it through the same output port (bunch) with probability
\begin{equation}
p_B=\frac{1+\langle \Pi_{ij}\rangle}{2}
\end{equation}
or remain in different modes (antibunch) with probability
\begin{equation}
p_A=\frac{1-\langle \Pi_{ij}\rangle}{2}
\end{equation}
This means that
\begin{equation}\label{2sym}
{\cal I}_{ij}=|p_B-p_A|^2.
\end{equation}

We would like to generalize the above operational measure to more than two particles. Before we start let us note  that in multipartite systems indistinguishability is related to all possible symmetries. Since in general permutation operators have only two common eigenvectors, it is natural to consider only the corresponding indistinguishable particles: totally symmetric bosons and totally antisymmetric fermions. However, one may also consider subgroups of permutation operators that define more general types of particle parastatistics. In the following we will investigate genuinely multipartite indistinguishability stemming from the subgroups generated by cyclic permutations.

The simplest example of such parastatistics can be obtained in a tripartite system. Consider the subgroup made of identity operator $\openone$, cyclic permutation $\Pi_{312}=\Pi_{23}\Pi_{12}$ and its inverse $\Pi_{231}=\Pi_{12}\Pi_{23}$. In analogy to Eq.(\ref{bim}), we define a measure of tripartite indistinguishability 
\begin{equation}\label{tim}
{\cal I}_{123} \equiv |\langle\psi|\Pi_{312}|\psi\rangle|^2 = |\langle\psi|\Pi_{231}|\psi\rangle|^2.
\end{equation}
The tripartite parastatistics correspond to states that maximize ${\cal I}_{123}$. To find them, we consider the eigenstates of cyclic permutation operator, which were also discussed by Peres in \cite{Peres}
\begin{eqnarray}
|\alpha\rangle &=& \frac{1}{\sqrt{3}} \left(|123\rangle + |312\rangle + |231\rangle \right), \label{alpha} \\
|\beta\rangle &=& \frac{1}{\sqrt{3}} \left(|123\rangle + \omega |312\rangle + \omega^2 |231\rangle \right), \\
|\gamma\rangle &=& \frac{1}{\sqrt{3}} \left(|123\rangle + \omega^2 |312\rangle + \omega |231\rangle \right), \\
|\bar{\alpha}\rangle &=& \frac{1}{\sqrt{3}} \left(|213\rangle + |132\rangle + |321\rangle \right), \\
|\bar{\beta}\rangle &=& \frac{1}{\sqrt{3}} \left(|213\rangle + \omega|132\rangle + \omega^2|321\rangle \right), \\
|\bar{\gamma}\rangle &=& \frac{1}{\sqrt{3}} \left(|213\rangle + \omega^2|132\rangle + \omega|321\rangle \right).\label{gamma}
\end{eqnarray} 
Here  $\omega = e^{i\frac{2\pi}{3}}$ and the notation $|xyz\rangle$ means that the first particle is in mode $x$, the second in mode $y$, and the third in mode $z$. For all these states ${\cal I}_{123}=1$, but ${\cal I}_{12}={\cal I}_{23}={\cal I}_{13}=0$, therefore they exhibit a genuinely tripartite indistinguishability without its bipartite counterpart. 

%%%%%%%%%%%%%%%%%%%%%%%%%%%%%%%%%%%%%%%%%%%%%%%%%%%%%%%%%%%%%%%%%%%%%%%%%%%%%%%%%%%%

{\it Symmetric operations.} Before we proceed, let us stress that studies on indistinguishability need to be based on symmetric operators. This is because application of asymmetric operators requires the ability to distinguish between the particles. However, we cannot a priori assume that the particles are distinguishable. This is what we want to check. Therefore,  as long as we do not gain access to a degree of freedom which can be used to effectively label and distinguish between the particles, we are fundamentally limited to symmetric operators.

Such operators commute with all permutation operators and therefore need to share their set of eigenvectors. However, in general permutation operators do not commute -- the only two common eigenstates they have are the symmetric state $|+\rangle$ and the antisymmetric state $|-\rangle$. Distinguishing between these two possibilities via symmetric operators is exactly the essence of the Hong-Ou-Mandel experiment.

On the other hand, none of the states given by Eqs,\;(\ref{alpha}-\ref{gamma}) is  a joint eigenvector of all permutation operators. In fact, bipartite permutation operators swap between the states $\{|\alpha\rangle,|\bar{\alpha}\rangle\}$, $\{|\beta\rangle,|\bar{\beta}\rangle\}$, and $\{|\gamma\rangle,|\bar{\gamma}\rangle\}$. This means that the states within each of these pairs cannot be distinguished using symmetric operators. However, one can check that the rank-2 states
\begin{eqnarray}
\rho_{\beta} &=& \frac{1}{2}(|\beta\rangle\langle\beta|+|\bar{\beta}\rangle\langle\bar{\beta}|), \\ 
\rho_{\gamma} &=& \frac{1}{2}(|\gamma\rangle\langle\gamma|+|\bar{\gamma}\rangle\langle\bar{\gamma}|), \\
\rho_{\alpha} &=& \frac{1}{2}(|\alpha\rangle\langle\alpha|+|\bar{\alpha}\rangle\langle\bar{\alpha}|)
\end{eqnarray}
commute with all the permutation operators. These three mixed states can be considered as representatives of the tripartite parastatistis. Note that $\text{Tr}\{\Pi_{312}\rho_{\alpha}\}=\text{Tr}\{\Pi_{312}\rho_{\beta}\}=\text{Tr}\{\Pi_{312}\rho_{\gamma}\}=1$, but $\text{Tr}\{\Pi_{ij}\rho_{\alpha}\}=\text{Tr}\{\Pi_{ij}\rho_{\beta}\}=\text{Tr}\{\Pi_{ij}\rho_{\gamma}\}=0$ for any $i\neq j$.  As a result, the problem of measuring the tripartite indistinguishability reduces to finding a method to perfectly distinguish between $\rho_{\alpha}$, $\rho_{\beta}$, and $\rho_{\gamma}$.

%%%%%%%%%%%%%%%%%%%%%%%%%%%%%%%%%%%%%%%%%%%%%%%%%%%%%%%%%%%%%%%%%%%%%%%%%%%%%%%%%%%%%%%%%%%%%%%%%%

{\it Detection of tripartite indistinguishability} As explained in the previous section, to experimentally measure ${\cal I}_{123}$ we should look for a simple natural process capable of perfectly discriminating between the states $\rho_{\alpha}$, $\rho_{\beta}$, and $\rho_{\gamma}$. In particular, we ask if it is possible to achieve it with free evolution, i.e., without interaction between the particles. We are going to show that the answer is positive.

Recall that each of the three particles can be in one of the three modes $|1\rangle$, $|2\rangle$, or $|3\rangle$. We consider a single-partite transformation between these modes given by the Quantum Fourier Transform (QFT) 
\begin{eqnarray}
U_{QFT}^{(3)} &=& \frac{1}{\sqrt{3}} \begin{pmatrix} 1 & 1 & 1 \\ 1 & \omega & \omega^2 \\ 1 & \omega^2 & \omega \end{pmatrix}. \label{tritter}
\end{eqnarray}
This transformation can be implemented  with a multiport commonly known as a {\it tritter} \cite{sciarrinotritter}. It can be visualized as a 3-port, i.e., a device with three inputs and three outputs. Since all the entries of $U_{QFT}^{(3)}$ have the same modulus, a single particle cast on the tritter is equally likely to end up in each of its output ports. In fact, a tritter can be represented by any unitary $3\times3$ matrix with this property, as they all generate equivalent dynamics.

Next, we apply $U=U_{QFT}^{(3)} \otimes U_{QFT}^{(3)} \otimes U_{QFT}^{(3)}$ to the states $\rho_{\alpha}$, $\rho_{\beta}$, $\rho_{\gamma}$. In other words, we feed the tritter with the tree particles that are in one of the above three states and observe the output ports. The tritter generates the following transformations on our basis states:
\begin{eqnarray}\label{evol}
& &|\alpha\rangle \rightarrow \frac{|111\rangle + |222\rangle + |333\rangle}{3} + \label{at} \\
& & \omega\frac{|213\rangle + |132\rangle + |321\rangle}{3} + \omega^2\frac{|123\rangle + |312\rangle + |231\rangle}{3}, \nonumber \\
& &|\beta\rangle \rightarrow \frac{|211\rangle + \omega |121\rangle + \omega^2 |112\rangle}{3} + \label{bt} \\
& & \frac{|322\rangle + \omega |232\rangle + \omega^2 |223\rangle}{3} + \frac{|133\rangle + \omega |313\rangle + \omega^2 |331\rangle}{3}, \nonumber \\
& &|\gamma\rangle \rightarrow \frac{|122\rangle + \omega |221\rangle + \omega^2 |212\rangle}{3} + \label{ct} \\
& & \frac{|233\rangle + \omega |332\rangle + \omega^2 |323\rangle}{3} + \frac{|311\rangle + \omega |113\rangle + \omega^2 |131\rangle}{3}. \nonumber
\end{eqnarray} 
Transformations on $|\bar{\alpha}\rangle$, $|\bar{\beta}\rangle$, and $|\bar{\gamma}\rangle$ are the same as on the unbarred states, with the only exception that one needs to swap $\omega \leftrightarrow \omega^2$. 

Let us focus on particle number measurements at the output ports of the tritter. We will denote their results as $\{n_1,n_2,n_3\}$, where $n_1+n_2+n_3 =3$ and $n_i$ is the number of particles detected at the $i^{\text{th}}$ port. The statistics of such measurements, presented in Tab. \ref{tab1}, offer a clear way of distinguishing between different  representatives of the tripartite parastatistics. If we detect only four possible particle number configurations: $\{1,1,1\}$, $\{3,0,0\}$, $\{0,3,0\}$ or $\{0,0,3\}$, we know that the corresponding state is a mixture of the totally symmetric and antisymmetric states (like $\rho_{\alpha}$) or their superposition. In case of $\{2,1,0\}$, $\{0,2,1\}$ or $\{1,0,2\}$ we can deduce that the state is $\rho_{\beta}$. Finally, the outcomes $\{1,2,0\}$, $\{0,1,2\}$, or $\{2,0,1\}$ indicate that the state is $\rho_{\gamma}$. For the last two cases we do not take into account superpositions in subspaces spanned by $\rho_{\beta}$ and $\rho_{\gamma}$. This is because, as we have already argued,  symmetric operations cannot distinguish between such superpositions.

\begin{table}[t]
  \begin{center}
    \caption{Probabilities of particle number counts after the tritter transformation on  three particles. We consider: three states corresponding to different parastatistics, symmetric state, antisymmetric state, state of three distinguishable particles and an arbitrary state $\rho$.}
    \label{tab1}
    \begin{tabular}{l|c|c|c|c|c|c|c} % <-- Alignments: 1st column left, 2nd middle and 3rd right, with vertical lines in between
      ~ & $\rho_{\alpha}$ & $\rho_{\beta}$ & $\rho_{\gamma}$ & $|+\rangle$ & $|-\rangle$ & $|123\rangle$ & $\rho$ \\
      \hline
$\{1,1,1\}$ & 2/3 & 0 & 0 & 1/3 & 1 & 2/9 & $p_{111}$ \\  
$\{3,0,0\}$ & 1/9 & 0 & 0 & 2/9 & 0 & 1/27 & $(p_{\alpha}-p_{111})/3$ \\
$\{0,3,0\}$ & 1/9 & 0 & 0 & 2/9 & 0 & 1/27 & $(p_{\alpha}-p_{111})/3$ \\
$\{0,0,3\}$ & 1/9 & 0 & 0 & 2/9 & 0 & 1/27 & $(p_{\alpha}-p_{111})/3$ \\
$\{2,1,0\}$ & 0 & 1/3 & 0 & 0 & 0 & 1/9 & $p_{\beta}/3$ \\
$\{0,2,1\}$ & 0 & 1/3 & 0 & 0 & 0 & 1/9 & $p_{\beta}/3$ \\
$\{1,0,2\}$ & 0 & 1/3 & 0 & 0 & 0 & 1/9 & $p_{\beta}/3$ \\
$\{1,2,0\}$ & 0 & 0 & 1/3 & 0 & 0 & 1/9 & $p_{\gamma}/3$ \\
$\{0,1,2\}$ & 0 & 0 & 1/3 & 0 & 0 & 1/9 & $p_{\gamma}/3$ \\
$\{2,0,1\}$ & 0 & 0 & 1/3 & 0 & 0 & 1/9 & $p_{\gamma}/3$ \\    
    \end{tabular}
  \end{center}
\end{table}

Since we consider a tripartite system in which each particle enters the tritter through a different port, a corresponding input state is a superposition, or a mixture, of the  six possible permutations of the state $|123\rangle$. Equivalently, an input state can be also represented in terms of the states given by  Eqs.\;(\ref{alpha}-\ref{gamma}). The formulas (\ref{at}), (\ref{bt}), and (\ref{ct})  show that their tritter transformations  lead to a particle count statistics that is describable by only four parameters (three, if one takes into account normalization). In particular, we observe that some events occur with the same probabilities
\begin{eqnarray}
p_{\{1,1,1\}} &\equiv& p_{111}, \\
p_{\{3,0,0\}} &=& p_{\{0,3,0\}} = p_{\{0,0,3\}} \equiv \frac{p_{\alpha}-p_{111}}{3}, \\
p_{\{2,1,0\}} &=& p_{\{0,2,1\}} = p_{\{1,0,2\}} \equiv \frac{p_{\beta}}{3}, \\
p_{\{1,2,0\}} &=& p_{\{0,1,2\}} = p_{\{2,0,1\}} \equiv \frac{p_{\gamma}}{3}.
\end{eqnarray}
Therefore, if an arbitrary state of three particles is fed into the tritter, the particle count statistics at the output is determined by $p_{\alpha}$, $p_{\beta}$, and $p_{\gamma}$ (see the last column in Tab. \ref{tab1}).  

The above observations allow us to arrive at the operational formula for the measure of tripartite indistinguishability (\ref{tim})
\begin{equation}
{\cal I}_{123} = \left|p_{\alpha} + \omega p_{\beta} + \omega^2 p_{\gamma}\right|^2.
\end{equation}
Note, that ${\cal I}_{123} = 1$ for $\rho_{\alpha}$, $\rho_{\beta}$, and $\rho_{\gamma}$. In addition, it is also equal to one for the tripartite symmetric and antisymmetric states. On the other hand, for a state of distinguishable particles, like $|123\rangle$ (see the fourth column in Tab. \ref{tab1}), we get $p_{\alpha}=p_{\beta}=p_{\gamma}=1/3$ and as a consequence ${\cal I}_{123} = 0$.

%%%%%%%%%%%%%%%%%%%%%%%%%%%%%%%%%%%%%%%%%%%%%%%%%

{\it N-partite indistinguishability.} So far, we have considered the simplest case of cyclic indistinguishability of three particles. Now we generalize our results to $n$-partite systems. Just as before, we define our measure of indistinguishability as the expectation value of the cyclic permutation operator
\begin{equation}\label{nim}
{\cal I}_{12\ldots n} \equiv |\langle\psi|\Pi_{n1\ldots (n-1)}|\psi\rangle|^2.
\end{equation}
Since $(\Pi_{n1\ldots (n-1)})^n=I$, the eigenvalues of the cyclic permutation operator are the $n^\text{th}$ roots of unity. They are all $(n-1)!$-degenerate and it is easy to verify that the vectors belonging to the same eigenspace can be converted into each other with a proper permutation of the particle labels. Because of that the eigenvectors corresponding to the same eigenvalue cannot be distinguished with symmetric operators. This means  that there are exactly $n$ indistinguishability classes stemming from ${\cal I}_{12\ldots n}$.

Let us represent these classes with the eigenvectors of form
\begin{multline}\label{nvec}
|\lambda^k\rangle=\frac{1}{\sqrt n}(|1,\;2,\,\ldots,\, n\rangle + \lambda^{k}|n,\,1,\,\ldots,\, n-1\rangle+\\
+\lambda^{2\,k}|n-1,\,n,\,\ldots,\, n-2\rangle+\ldots+\lambda^{(n-1)k}|2,\,3,\, \ldots,\, n,\,1\rangle),
\end{multline}
where $k\in\{1,\ldots,n\}$ and $\lambda= e^{i\frac{2\pi}{n}}$. For all these states the $n$-partite indistinguishability measure $\mathcal{I}_{1,2,\ldots,n} = \bra{\psi} \Pi_{n,1,\ldots,(n-1)} \ket{\psi} = 1$ while $\mathcal{I}_{1,2,\ldots,m} = \bra{\psi} \Pi_{m,1,\ldots,(m-1)} \ket{\psi} = 0$ for all $m < n$. This means they are genuinely $n$-partite indistinguishable. 

In order to find an operational formula for ${\cal I}_{12\ldots n}$, we study the evolution of the states $|\lambda^k\rangle$ on a Fourier multiport given by
\begin{equation}
U_{QFT}^{(n)}|j\rangle = \frac{1}{\sqrt{n}}\sum_{k=1}^{n} e^{i\frac{2\pi}{n}(j-1)(k-1)}|k\rangle. \label{QFT} \\
\end{equation}
Just as in the tripartite case we look for the outputs that are characteristic for each indistinguishability class. This time, however, we will use a suppression law proposed in  \cite{Dittel18}. It links the set of allowed output events with the symmetry of the input state and its transformation. Applied to the evolution of the state $|\lambda^k\rangle$ it yields the condition

\begin{equation}\label{con}
\sum_{i=0}^{n-1} i a_i  \equiv k \;(\text{mod}\;n)
\end{equation}
where $a_0, a_1,\ldots$ denote the number of particles  in each consecutive output port. Clearly, if a specific configuration $(a_0,a_1,\ldots,a_{n-1})$ solves the equation for $\lambda^k$, it cannot solve it for any other  eigenvalue $\lambda^j$. This means that the sets
\begin{equation}
A_k:=\{(a_0,a_1,\ldots,a_{n-1}) : \sum_{i=0}^{n-1} i a_i \equiv k \;(\text{mod}\;n)\},
\end{equation}
consisting of outputs stemming from the evolution of states of different indistinguishability classes are completely disjoint. 
Thus we can define 
\begin{equation}
p_{k}:=\sum_{ (a_0,a_1,\ldots,a_{n-1}) \in A_k} p_{\{a_0,a_1,\ldots,a_{n-1}\}}
\end{equation}
and use it to obtain the operational formula for  ${\cal I}_{12\ldots n}$

\begin{equation}
\mathcal{I}_{1,2,\ldots,m}=|\sum_{i=0}^{n-1} p_i \lambda^i|^2.
\end{equation}

%%%%%%%%%%%%%%%%%%%%%%%%%%%%%%%%%%%%%%%%%%%%%%%%%%%%%%%%%%%%%%%%%%%%%%%%%%%%%%%%

{\it Summary.} We have shown that the concept of genuine multipartite indistinguishability naturally emerges if we define indistinguishability with respect to cyclic permutations. Then, we found that this new type of indistinguishability can be tested using symmetric multiports in setups that generalize the Hong-Ou-Mandel one. We believe that our results can find applications in various fields of quantum information science, such as resource theories, quantum walks, secret sharing protocols and the description of multipartite correlations.  

%%%%%%%%%%%%%%%%%%%%%%%%%%%%%%%%%%%%%%%%%%%%%%%%%%%%%%%%%%%%%%%%%%%%%%%%%%%%%%%%

{\it Acknowledgements.} This research was supported by the National Science Center in Poland. P.K. acknowledges NCN Grant No. 2014/14/E/ST2/00585. M.K. acknowledges NCN Grant No. 2014/14/E/ST2/00585 (the initial stage of research) and NCN Grant No. 2017/27/N/ST2/01858 and the doctoral scholarship no. 2018/28/T/ST2/00148 (the final stage of research). R.P. is supported by EPSRC studentship (UK).

%%%%%%%%%%%%%%%%%%%%%%%%%%%%%%%%%%%%%%%%%%%%%%%%%%%%%%%%%%%%%%%%%%%%

\bibliographystyle{apsrev}

%\bibliography{indist_biblio} 

\begin{thebibliography}{19}
\bibitem[{not()}]{note}
\bibinfo{note}{The reader may be confused that we have omitted anyons
  \cite{leinaas77}, which can be observed in some special conditions. This is
  because here we focus on the properties of the permutation group, which
  describes the interchange of particles in the three-dimensional space. On the
  other hand, anyons are in fact complex particles in two-dimensional space
  whose interchange is described by the braid group.}
\end{thebibliography}

%%%%%%%%%%%%%%%%%%%%%%%%%%%%%%%%%%%%%%%%%%%%%%%%%%%%%%%%%%%%%%%%%%%%%%%%%%%%%%%%%%%%%%%%

\end{document}